\documentclass{article}
\usepackage[T1]{fontenc} % add special characters (e.g., umlaute)
\usepackage[utf8]{inputenc} % set utf-8 as default input encoding
\usepackage{ismir,amsmath,cite,url}
\usepackage{graphicx}
\usepackage{color}
\usepackage{booktabs}
\usepackage{multirow}
 \usepackage{hhline}
\usepackage{xcolor}
\usepackage{adjustbox}
\usepackage{rotating}
\usepackage{tablefootnote}
\usepackage{threeparttable}
\usepackage{enumitem}
\usepackage{kotex}
\usepackage{comment}
\usepackage{colortbl}

\usepackage{lineno}

\title{LP-M\lowercase{usic}C\lowercase{aps}: LLM-Based Pseudo Music Captioning}

\multauthor
{SeungHeon Doh$^{\flat}$ \hspace{1cm} Keunwoo Choi$^{\natural}$ \hspace{1cm} Jongpil Lee$^{\sharp}$ \hspace{1cm} Juhan Nam$^{\flat}$} {
 $^{\flat}$ Graduate School of Culture Technology, KAIST, South Korea \\
$^{\natural}$ Gaudio Lab, Inc., South Korea\\
$^{\sharp}$ Neutune, South Korea\\
{\tt\small \{seungheondoh, juhan.nam\}@kaist.ac.kr, keunwoo@gaudiolab.com, jongpillee@neutune.com}
}

\def\authorname{SeungHeon Doh, Keunwoo Choi, Jongpil Lee, Juhan Nam}

\begin{document}

\maketitle
\begin{abstract}
Automatic music captioning, which generates natural language descriptions for given music tracks, holds significant potential for enhancing the understanding and organization of large volumes of musical data. Despite its importance, researchers face challenges due to the costly and time-consuming collection process of existing music-language datasets, which are limited in size. 
To address this data scarcity issue, we propose the use of large language models (LLMs) to artificially generate the description sentences from large-scale tag datasets. This results in approximately 2.2M captions paired with 0.5M audio clips. We term it \textbf{L}arge Language Model based \textbf{P}seudo music caption dataset, shortly, \textbf{LP-MusicCaps}. We conduct a systemic evaluation of the large-scale music captioning dataset with various quantitative evaluation metrics used in the field of natural language processing as well as human evaluation. In addition, we trained a transformer-based music captioning model with the dataset and evaluated it under zero-shot and transfer-learning settings. The results demonstrate that our proposed approach outperforms the supervised baseline model.\footnote{Our dataset and codes are available at \url{https://github.com/seungheondoh/lp-music-caps}}
% In addition, we train a transformer-based music captioning model with the dataset in a supervised, zero-shot, and transfer-learning setting, and show that it outperforms the baseline models.

%Automatic music captioning plays a crucial role in the field of music information retrieval as it facilitates the comprehension and organization of vast amounts of musical data by generating natural language descriptions. To address this data scarcity issue, we propose the use of tag-to-caption augmentation with large language models (LLMs). We also propose a systemic evaluation of the LLM-based augmentation for the first time. Our approach results in the creation of \textbf{LP-MusicCaps}, a \textbf{L}arge Langauge Model based \textbf{P}seudo music caption dataset. With three existing dataset (MusicCaps, MSD, MTT), LP-MusicCaps consist of approximately 0.5M audio clips paired with 2M captions. Using the LP-MusicCaps dataset, we train a music captioning model which outperforms the existing baseline and shows improved generalized. \footnote{We uploaded more examples from the LP-MusicCaps dataset at this anonymous site: \url{https://sites.google.com/view/ismir23-lp-music-caps}}
\end{abstract}
\section{Introduction}\label{sec:introduction}
Music captioning is a music information retrieval (MIR) task of generating  natural language descriptions of given music tracks. The text descriptions are usually sentences, distinguishing the task from other music semantic understanding tasks such as music tagging. Recently, there have been some progress in music captioning including track-level captioning~\cite{manco2021muscaps, cai2020music} and playlist-level captioning~\cite{choi2016towards, doh2021music, gabbolini2022data, kim2023music}.
These approaches usually utilize a deep encoder-decoder framework which is originally developed for neural machine translation \cite{bahdanau2014neural}. 
Choi~\textit{et~al.}\cite{choi2016towards} used a pre-trained music tagging model as a music encoder and an RNN layer initialized with pre-trained word embeddings for text generation. 
%However, the model fails to generate sentence, mainly due to overfitting on a small training set. More recently, 
Manco~\textit{et~al.}~\cite{manco2021muscaps} introduced a temporal attention mechanism for alignment between audio and text by pairing a pre-trained harmonic CNN encoder~\cite{won2020data} with an LSTM layer.  
% They utilized not only a pre-trained audio encoder but also a pre-trained text decoder to address the issue of insufficient data.  
%To address the issue of insufficient data, the authors utilized not only a pre-trained audio encoder but also a pre-trained text decoder.
Gabbolini~\textit{et~al.}~\cite{gabbolini2022data} generated playlist titles and descriptions using pre-trained GPT-2~\cite{brown2020language}.% However, due to the issue of data scarcity, previous works have been unable to adopt an end-to-end approach.

Currently, the primary challenge of track-level music captioning is the scarcity of large-scale public datasets.  Manco~\textit{et~al}.\cite{manco2021muscaps} used private production music datasets. 
% attempted to build a music captioning model using a private production music dataset consisting of 6053 audio-caption pairs. 
% However, neither the pre-train model nor the dataset has been publicly disclosed.
Huang~\textit{et~al.}\cite{huang2022mulan} also used a private dataset with 44M music-text pairs on YouTube, but this approach is hardly reproducible or affordable for other researchers. 
To address this data issue, a community-driven data collection initiative has been proposed \cite{manco2022song}.
% \footnote{It has not yet been made publicly available as of early 2023.}
As of now, the only publicly available dataset for track-level music captioning is MusicCaps~\cite{agostinelli2023musiclm}, which includes high-quality music descriptions from ten musicians. However, it is limited to 5521 music-caption pairs as it was originally created as an evaluation set for a text-prompt music generator.

% The main bottleneck of music captioning is the lack of large-scale dataset. To address this issue, a community-driven data collection was proposed \cite{manco2022song}. A large-scale data crawling resulted in a music captioning dataset \cite{agostinelli2023musiclm}, although the actual publicized dataset only consists of 5,521 tracks. As alternatives, Kim~et~al. assumed playlist titles are equivalent to music captions \cite{kim2023music}, (maybe Todo;Doh ?? et al., did this, and that, .. etc.)

\begin{figure}[!t]
\centering
\includegraphics[width=0.85\linewidth]{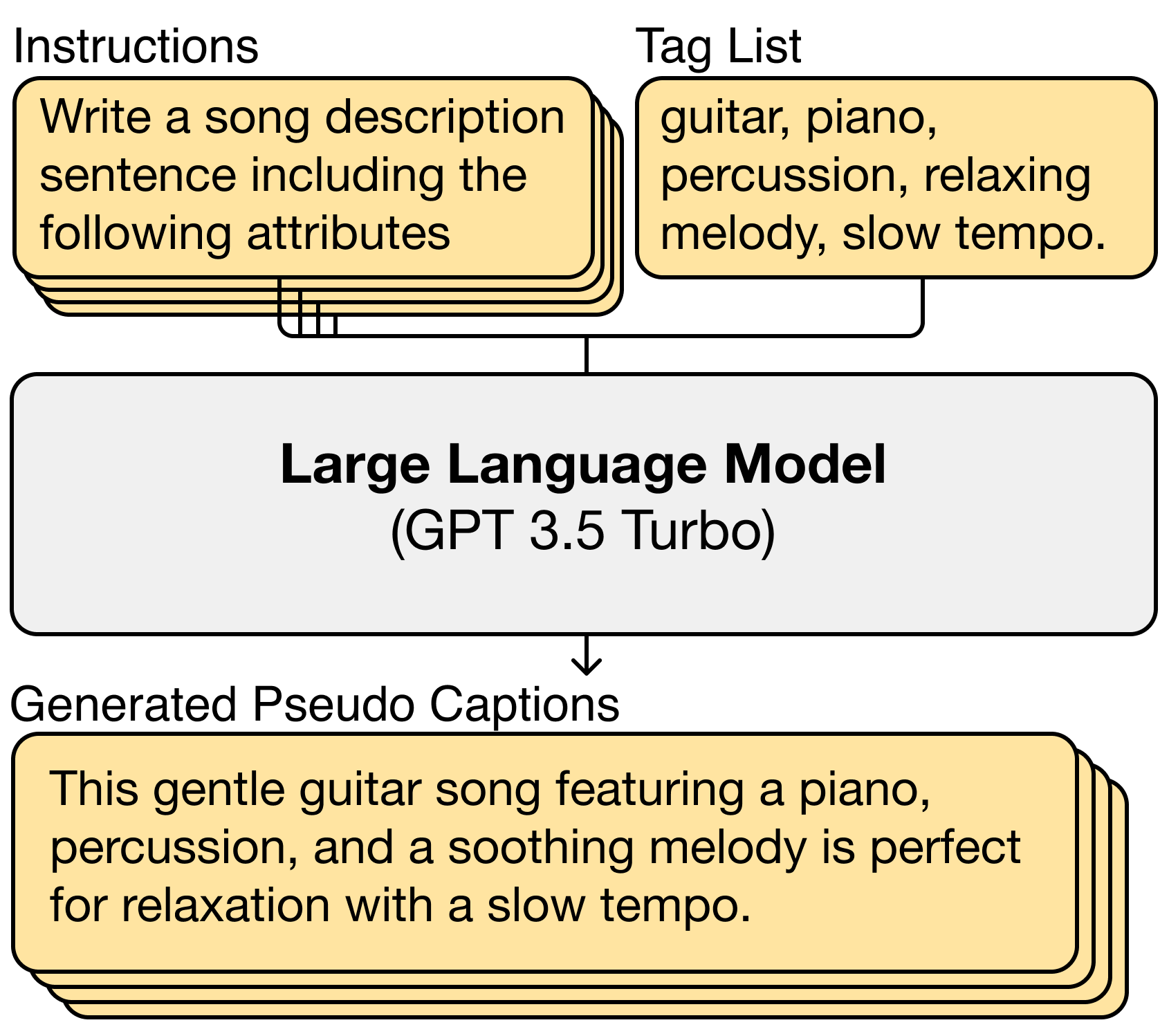}
\vspace{-3mm}
\caption{The generation process of pseudo captions by feeding a large language model with instructions and manually-annotated labels.}
\vspace{-3mm}
\end{figure}

With the scale of the aforementioned datasets, it remains difficult to train a music captioning model successfully. A workaround for this situation is to use music tagging datasets and generate sentences with tag concatenation~\cite{cai2020music,doh2022toward} or prompt template~\cite{chen2022learning}. As relying on tagging datasets, however, the tag-to-sentence approaches would have the same limitation tagging datasets have. For example, high false-negative rates of tagging datasets \cite{choi2017effects}. Tagging datasets also has some typical issues text data have, for example, synonyms, punctuation, and singular/plural inconsistencies. Without proper treatment, these can limit the performance of the corresponding music captioning models. 

A potential solution is to use strong language models, i.e., large language models (LLMs). LLMs refer to the recent large-scale models with over a billion parameters that exhibit strong few-shot and zero-shot performance  \cite{10.5555/3455716.3455856, brown2020language}. Large language models are usually trained with text data from various domains such as Wikipedia, GitHub, chat logs, medical articles, law articles, books, and crawled web pages~\cite{gao2020pile}. When successfully trained, they demonstrate an understanding of words in various domains \cite{brown2020language}. There have been similar and successful use cases of LLMs for general audio understanding~\cite{mei2023wavcaps} and music generation~\cite{huang2023noise2music}.

Motivated by the recent success of LLMs, we propose creating a music captioning dataset by applying LLMs carefully to tagging datasets. Our goal is to obtain captions that are i)~semantically consistent with the provided tags, ii)~grammatically correct, and iii)~with clean and enriched vocabulary. This dataset-level approach is rather pragmatic than sophisticated; it alleviates the difficulty of music captioning tasks not by theory or model, but by data. The aforementioned ambiguous aspects of the music captioning task are addressed by the powerful LLMs that cost reasonably~\cite{gilardi2023chatgpt}, considering the training cost music researchers would spend otherwise. Once the creation is complete, it is straightforward to train some music captioning models by supervised learning. 

% Motivated by the recent success of LLMs, we propose creation of music captioning dataset. Our proposal is to apply LLMs carefully to tagging datasets to have music-caption pairs. 

There are some existing works in the pseudo-labeling using language models. Huang~\textit{et~al.}~\cite{huang2023noise2music} introduced the MuLaMCap dataset, which consists of 400k music-caption pairs generated using the large language model and the music-language joint embedding model. They utilized a large language model (LaMDA\cite{thoppilan2022lamda}) to generate 4M sentences using 150k song metadata as input in the format of \texttt{\{title\} by \{artist\}}. Then the text and music-audio joint embedding model, MuLan, calculates the similarity between music and generated captions, annotating pairs with high similarity \cite{huang2022mulan}. However, it is not possible to reproduce or evaluate this work as the adopted language model as well as the final music-audio embedding model are not publicly available. Moreover, using metadata has some issues
-- a popularity-biased, limited coverage and a low reliability --
as we discuss later in Section~\ref{sec:LLM_data_gen}. 
% \jp{여기에서 song metadata는 객관적인 정보의 학습에 안 좋은 영향이 있을 수 있다고 언급해주는건 어떨지 궁금해요, (2.1의 마지막에서 2번째 문장을 읽다가, song metadata가 hallucination 효과를 일으킨다는게 앞에서 언급이 없었던거 같아서요.)} 
% keunwoo: 좋습니다. 위에 써있는대로 (-- a popularity-biased, limited coverage and a low reliability --) 추가했습니다. 
Wu~\textit{et~al.}\cite{wu2022large} introduce keyword-to-caption augmentation (K2C Aug) to generate captions based on the ground truth tags of audio clips in AudioSet. They used a pre-trained T5 model without any instruction. Finally, Mel~\textit{et~al.}.~\cite{mei2023wavcaps} introduce WavCaps, a 400k audio captioning dataset using ChatGPT~\cite{ouyang2022training}. However, previous approaches only reported task performance and did not directly evaluate the quality of generated captions.
% sh: 공간 문제로 빼겠습니다! \footnote{https://github.com/gagan3012/keytotext}

We propose a solution in this paper with three-fold key contribution. First, we propose an LLM-based approach to generate a music captioning dataset, \textbf{LP-MusicCaps}. Second, we propose a systemic evaluation scheme for music captions generated by LLMs. Third, we demonstrate that models trained on LP-MusicCaps perform well in both zero-shot and transfer learning scenarios, justifying the use of LLM-based pseudo-music captions.

\begin{table*}[]
\resizebox{\textwidth}{!}{
\begin{tabular}{l}
\toprule
{\color[HTML]{9B9B9B} Input tags} \\
video game theme, no singer, instrumental, analog sounding, small keyboard, beatboxing, playful, cheerful, groovy \\ \midrule
\midrule
{\color[HTML]{9B9B9B} Result of "Writing" Instruction} \\ 
\begin{tabular}[c]{@{}l@{}}This instrumental track has a joyful and playful vibe, perfect for a video game theme. With no singer, the analog-sounding music features a \\ small keyboard and beatboxing, creating a groovy and cheerful atmosphere.\end{tabular} \\ \midrule
{\color[HTML]{9B9B9B} Result of "Summary" Instruction} \\ 
A cheerful and groovy instrumental track with playful beatboxing and an analog sound, featuring a small keyboard and video game-inspired \\ theme. \\ \midrule
{\color[HTML]{9B9B9B} Result of "Paraphrase" Instruction} \\ 
\begin{tabular}[c]{@{}l@{}}This upbeat instrumental tune captures the essence of a fun-filled video game theme, incorporating analog-sounding beats,\\ playful beatboxing rhythms, and a groovy small keyboard melody, all without the need for a singer.\end{tabular} \\ \midrule
{\color[HTML]{9B9B9B} Result of "Attribute Prediction" Instruction} \\
\begin{tabular}[c]{@{}l@{}}\{"new\_attribute": {[}"8-bit sound", "chiptune style", "retro vibe"{]},\\ "description": "This instrumental tune is straight out of a video game with its analog sounding melodies and small keyboard tinkles. \\ Beatboxing adds a playful element to the groovy, cheerful vibe. Reminiscent of classic 8-bit sound and chiptune style, this retro vibe\\ is sure to put a smile on your face."\}\end{tabular} \\ \midrule
\midrule
{\color[HTML]{9B9B9B} Ground Truth} \\
\begin{tabular}[c]{@{}l@{}}This is a video game theme cover. The theme belongs to the Super Mario franchise. The main theme is being played on an analog sounding \\small keyboard. There is an added rhythmic background of beatboxing in this version. The atmosphere is playful. This piece could be used \\ in the background of arcade gaming social media content.\end{tabular} \\ \bottomrule
\end{tabular}
}
\vspace{-3mm}
\caption{An example of generated captions from MusicCaps dataset.}
\vspace{-3mm}
\label{tab1:example}
\end{table*}

\section{Pseudo Caption Generation using Large Language Models}\label{sec:caption_gen}
In this section, we introduce how music-specific pseudo captions are created using a large language model in the proposed method. 

\subsection{Large Language Model for Data Generation}
\label{sec:LLM_data_gen}
We first take multi-label tags from existing music tagging datasets. The list of tags are appended with a carefully written task instruction as an input (prompt) to a large language model. The model then generates and returns sentences that (may) describe the music in a way the task instruction conditions. Table~\ref{tab1:example} shows examples of generated captions according to multi-label tags and task instructions. For the language model, we choose GPT-3.5~Turbo~\cite{ouyang2022training} for its strong performance in various tasks. During its training, it was first trained with a large corpus and immense computing power, then fine-tuned by reinforcement learning
with human feedback (RLHF)~\cite{christiano2017deep} for better interaction with given instruction. As a result, GPT-3.5~Turbo demonstrates state-of-the-art zero-shot abilities in understanding, reasoning, and generating human-like responses to natural language inputs. 

% keunwoo: LLMs을 사용할때 주의해야할 점 소개. "hallucination" 소개, 설명, 인용. 많은 deep learning 모델과 마찬가지로 LLMs도 출력의 confidence를 측정하기가 어려우며, 이런 점이 creative application에서는 도움이 되지만 우리에겐 도움이 안됨 왜냐면 우린 factual해야 하므로. 그리고 우리는 LLM을 naive하게 사용하지 않는다는점 강조. preliminary experiment에서 가짜 정보가 추가되지 않도록 확인하고, 이를 subjective evaluation 에도 집어넣어서 확인했다는 것을 강조.
Since LLMs contain a wide range of information, music captions may be generated based on some famous musical entities such as the artist name or album name. However, LLMs may generate inaccurate text in a confident tone which is hard to detect without ground truth. This issue, known as hallucination, can be a fun aspect when using LLMs for creative purposes~\cite{Ji_2023}. However, hallucination should be avoided in an application like ours as the resulting captions should be factual. Therefore, we do not use any metadata unlike a previous work \cite{huang2023noise2music}. We also added a question to measure hallucination in the proposed evaluation scheme. 
 % , we do not use metadata as input because low-quality captions or hallucinations may be generated for newly released or unpopular tracks. % 부연설명. LLM이 어떤 musical entity (artist, album, track)에 대해 정확한 정보를 알고 있는지 아닌지를 알기 어려움. 
% Moreover, there is no open-source text-music audio joint embedding model available to prevent this issue.

\subsection{Task Instruction Design}\label{sec:instruction}
Our proposed caption generation follows the formulation: $\Tilde{y}_{\text{cap}} = f_{\text{LLM}}(y_{\text{tag}}, i)$, where $y_{\text{tag}}$ and $\Tilde{y}_{\text{cap}}$ refer to the multi-label tag and the generated caption, respectively, and $i$ is the task instruction provided. Given that the output can vary based on the task instruction, even with the same model and input, task instructions become a crucial aspect of data generation. Therefore, we define four different tasks and generate captions accordingly.

\vspace{2mm}
\noindent \textbf{Writing}: \textit{Write a song description sentence including the following attributes.}  \texttt{\{input tags\}}

\vspace{2mm}
\noindent \textbf{Summary}: \textit{Write a single sentence that summarizes a song with the following attributes. Don't write the artist name or album name. }\texttt{\{input tags\}}

\vspace{2mm}
\noindent \textbf{Paraphrase}: \textit{Write a song description sentence including the following attributes. Creative paraphrasing is acceptable.} \texttt{\{input tags\}}

\vspace{2mm}
\noindent \textbf{Attribute Prediction}: \textit{Write the answer as a Python dictionary with new\_attribute and description as keys. For new\_attribute, write new attributes that show high co-occurrence with the following attributes. For description, write a song description sentence including the following attributes and new attributes.} \texttt{\{input tags\}}

\begin{table*}[]
\resizebox{\textwidth}{!}{
\begin{tabular}{lcclccccccclcclc}
\toprule
 &  &  &  & \multicolumn{7}{c}{Supervised Metrics} &  & \multicolumn{2}{c}{Diversity Metrics} &  & Length \\ \cmidrule{5-11} \cmidrule{13-14} \cmidrule{16-16} 
Methods & LM & Params &  & B1$\uparrow$ & B2$\uparrow$ & B3$\uparrow$ & B4$\uparrow$ & M$\uparrow$ & R-L$\uparrow$ & BERT-S$\uparrow$ &  & Vocab$\uparrow$ & $\text{Novel}_{\text{v}}\uparrow$ &  & Avg.Token \\ \midrule
\multicolumn{16}{l}{{\color[HTML]{9B9B9B} Baseline}} \\
Tag Concat~\cite{cai2020music,doh2022toward} & - & - &  & 20.25 & 13.57 & 8.64 & 5.42 & 23.24 & 19.52 & 86.24 &  & 3506 & 46.92 &  & 20.6\small{±11.2} \\
Template~\cite{chen2022learning} & - & - &  & 25.41 & 16.15 & 10.00 & 6.15 & 25.57 & 21.36 &  87.92 &  & 3507 & 46.93 &  & 25.6\small{±11.2} \\
K2C Aug.~\cite{wu2022large} & T5 & 220M &  & 6.07 & 3.01 & 1.58 & 0.85 & 14.23 & 17.92 &  86.33 &  & 3760 & \textbf{67.66} &  & 14.7\small{±5.1} \\ \midrule
\multicolumn{16}{l}{{\color[HTML]{9B9B9B} Proposed Instruction}}  \\
Writing & GPT3.5 & 175B+ &  & \textbf{36.84} & \textbf{19.85} & \textbf{11.37} & \textbf{6.74} & 31.44 & 25.36 &  89.26 &  & 5521 & 56.17 &  & 44.4\small{±17.3} \\
Summary & GPT3.5 & 175B+ &  & 26.12 & 14.58 & 8.80 & 5.52 & 27.58 & \textbf{25.83} &  \textbf{89.88} &  & 4198 & 49.52 &  & 28.6\small{±10.7} \\
Pharapase & GPT3.5 & 175B+ &  & 36.51 & 18.73 & 10.33 & 5.87 & 30.36 & 23.40 & 88.71 &  & 6165 & 59.95 &  & 47.9\small{±18.7} \\
Attribute Prediction  & GPT3.5 & 175B+ &  & 35.26 & 18.16 & 9.69 & 5.41 & \textbf{34.09} & 23.19 & 88.56 &  & \textbf{6995} & 63.16 &  & 66.2\small{±21.6}
\\ \bottomrule
\end{tabular}
}
\vspace{-3mm}
\caption{Performance of existing pseudo caption generation methods and the proposed method. LM stand for the language model. Avg.Token stand for the average number of token per caption.}
\vspace{-1mm}
\label{tab2:nlp-metrics}
\end{table*}

\begin{figure*}[!t]
\centering
\includegraphics[width=0.96\textwidth]{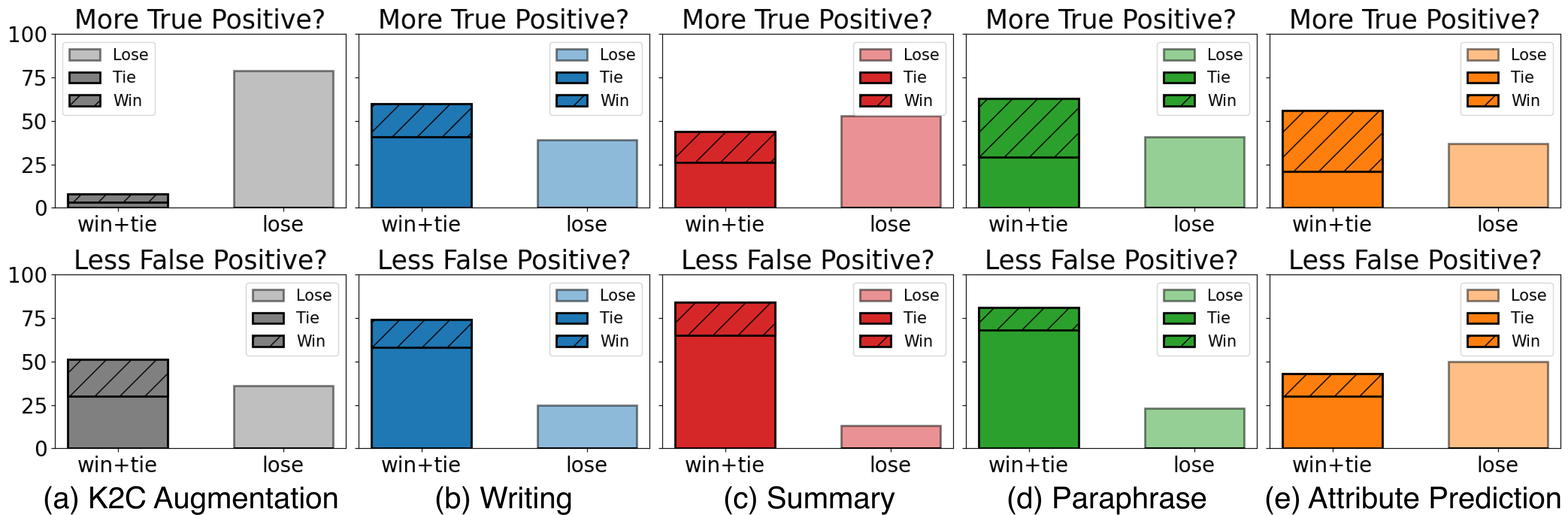}
\vspace{-1.5mm}
\caption{A-vs-B test results. Each method is compared to ground truth in terms of having more true positives and fewer false positives. The proposed methods (b, c, d, e) show comparable \textbf{win+tie} performance to ground truth.}
\vspace{-3mm}
\label{fig2:ab-test}
\end{figure*}

\vspace{2mm}
In every instruction, we add `include / with the following attributes' to prevent hallucination. The ``Writing'' task instruction is a simple prompt that uses tags to generate a sentence.  The ``Summary'' task instruction aims to compress information into a short length. The ``Paraphrase'' task instruction expands the vocabulary. Finally, the ``Attribute Prediction'' task instruction predicts new tags based on tag co-occurrence in large corpora (i.e. the training data of GPT-3.5~Turbo), which is expected to address the issue of high false-negative rates in existing tagging datasets while mitigating the risk of hallucination. In this instruction, `new attributes' exists to bridge the description and the input, and we only use the `description' as caption. 
% \sh{To identify predict attributes, we request 
% we request both the new attributes and song description written with input tags and new attributes, we only use the song description during training models.}

\section{Evaluation of Pseudo Captions}
% Despite obtaining competitive generated results,  % keunwoo: 아니 competitive한지 어떻게 알아요 
% We conduct evaluation to address the hallucination issue. We also evaluate the semantic correspondence between music and generated captions through user tests.  % keunwoo: 이 문장 일단 좀 어색하기도하고, 여기에서 굳이 뭐 말해야하나요? 
It is crucial to ensure the quality of generated captions, especially since they are supposed to be used as ground truth. In this section, we introduce a holistic evaluation scheme that includes objective and subjective assessment -- and its result on the captions from the proposed method. 

\subsection{Objective Evaluation}\label{sec:obj}
We conduct evaluation on the generated captions using MusicCaps dataset \cite{agostinelli2023musiclm}. It has audio ($x$), tag list ($y_{\text{tag}}$), and ground truth caption ($y_{\text{cap}}$). The pseudo captions ($\Tilde{y}_{\text{cap}}$) are generated with four pre-defined instructions as explained in Section \ref{sec:instruction} for all items in the evaluation split. During the evaluation, the generated captions are compared to the ground truth captions with respect to \textit{n}-gram, neural metrics. We also report diversity metrics.

% During the evaluation, the generated captions are compared to the ground truth captions with respect to \textit{n}-gram, neural, and diversity metrics. 

Following the previous work~\cite{gabbolini2022data}, we measure four \textit{n}-gram metrics~\cite{papineni2002bleu, banerjee2005meteor, lin2004rouge}: BLEU1 to 4 (B1, B2, B3, B4), METEOR (M), and ROUGE-L (R-L). They are all based on \textit{n}-gram precision and recall between the ground truth and generated captions. These metrics capture different aspects of the caption quality. BLEU and METEOR focus on \textit{n}-gram overlap between the generated and ground truth captions, while ROUGE-L measures the longest common sub-sequence between the two. 

% Following the previous work~\cite{gabbolini2022data}, we measure four \textit{n}-gram metrics~\cite{papineni2002bleu, banerjee2005meteor, lin2004rouge, vedantam2015cider}: BLEU1 to 4 (B1, B2, B3, B4), METEOR (M), ROUGE-L (R-L), and CIDEr (C). They are all based on \textit{n}-gram precision and recall between the ground truth and generated captions. These metrics capture different aspects of the caption quality. BLEU and METEOR focus on \textit{n}-gram overlap between the generated and ground truth captions, while ROUGE-L measures the longest common sub-sequence between the two. CIDEr, on the other hand, considers the importance of rare words and the word frequency in the caption by applying TF-IDF weighting.

In addition, we use BERT-Score (BERT-S) based on pre-trained BERT embeddings to represent and match the tokens in the ground truth with respect to the generated caption \cite{zhang2019bertscore}. By computing the similarity between the BERT embeddings of each token, BERT-Score can better capture the semantic similarity between the generated and ground truth captions than \textit{n}-gram metrics; as it is more robust to synonyms, paraphrasing, and word order variations.

Finally, we evaluate the diversity of the generated captions by measuring how many different words are used. \textit{novel$_{v}$} indicates the percentage of new vocabulary in generated captions that are not among the training vocabulary. \textit{Vocab} is the number of unique words used in all the generated captions. It is worth noting that diversity metrics are generally considered as subsidiaries and do not capture the overall quality of the generated captions.

% \footnote{Unlike \cite{gabbolini2022data}, we do not measure the percent of unique captions because all of the generated captions are novel.} 
% sh: 이후 music captioning에서 포함하기도 하고, 공간문제로 뻈습니다!

\subsection{Subjective Evaluation}
Following the previous work~\cite{agostinelli2023musiclm}, we set up an A-vs-B human rating task, in which a participant is presented with a 10-second single music clip and two text descriptions. We randomly selected 240 music samples from the MusicCaps evaluation dataset. Since the research goal is to generate music captions that can be used as pseudo-ground truth, one description is always fixed to the ground truth and the other is chosen from 5 types of generated captions including the K2C Augmentation~\cite{wu2022large} and the four proposed instruction methods. This yields up to 1200 (= 240 x 5) questions. We hired 24 participants who are music researchers or professionals in the music industry. Each of them rated 20 randomly selected questions. As a result, we collected a total of 480 ratings. 
%Out of a total of 1200 (= 240 x 5) questions, each participant who is a music researcher or professional in the music industry rated 20 random questions.
% Since the research goal is to generate music captions that can be used as pseudo-ground truth, one description is always the ground truth and the other four are generated captions from the five different methods. Out of a total of 1200 pairs, each participant was asked to answer 20 questions (i.e., 20 random pairs). 
The rater was asked to evaluate caption quality on two different aspects: (Q1) \textit{More True Positive}: which caption describes the music with more accurate attributes? (Q2) \textit{Less False Positive}: which caption describes the music less wrong? For example, if a method produces long and diverse sentences with many music attributes, it may be advantageous for Q1 but disadvantageous for Q2. Conversely, if a method conservatively produces short sentences with few music attributes, it may be advantageous for Q2 but disadvantageous for Q1. We determine the ranking of conditions by counting the number of wins, ties, and loses in the pairwise tests.
% Since this is a subjective test, each user may have different ground truths, and generated samples may sometimes outperform ground truth.

\subsection{Results}
We compare our LLM-based caption generation with two template-based methods (tag concatenation, prompt template\footnote{Template example: the music is characterized by \texttt{\{input tags\}}}) and K2C augmentation \cite{wu2022large}. In Table \ref{tab2:nlp-metrics}, we present the captioning result for MusicCaps~\cite{agostinelli2023musiclm} evaluation set. When comparing our proposed method with existing methods, we observe significant differences in \textit{n}-gram metrics. This is because the tag concatenation fails to complete the sentence structure. In the case of K2C Augmentation, due to the absence of instruction, the input tag is excluded from the generated caption, or a sentence unrelated to the song description sentence is created. In contrast, the template-based model shows improved performance as the musical context exists in the template. We next consider diversity metric with BERT-Score. Our proposed method shows higher values in BERT-Score while generating diverse vocabularies. This indicates that the newly created vocabulary does not harm the music semantics.

Comparing within the proposed different task instructions, we can observe that each instruction performs a different role. ``Writing'' shows a high \text{n}-gram performance as it faithfully uses input tags to generate captions. ``Summary'' has the smallest average number of tokens due to its compression of information, but it shows competitive performance in ROUGE-L which is specialized to summarizing, as well as the highest BERT-Score. ``Paraphrase'' generates many synonyms, resulting in a large vocabulary size and the use of novel vocabulary. ``Attribute Prediction'' predicts new tags based on the co-occurrence of tags. This instruction shows lower performance in BLEU but competitive results in METEOR, which utilizes a thesaurus, such as WordNet, to consider the accuracy scores of words with similar meanings, indicating that newly predicted tags have similar semantic with ground truth. 

% 제안된 서로다른 instruction을 비교할떄, 우리는 각 instruction이 서로다른 역활을 수행하는것을 볼수 있습니다. wrtie는 input tag를 충실하게 사용하여 글을 작성하기 떄문에 높은 N-Gram performance를 가집니다. 반면 summary는 정보를 압축하기 떄문에 가장 작은 average number of token을 가지지만 높은 Bert Score를 가지게 됩니다. Pharapase는 많은 synonym을 생성하는 task입니다. 때문에 높은 vocab size와 novel vocab을 사용하게 됩니다. 마지막으로 Predict Attribute는 태그의 동시등장성을 기반으로 새로운 태그를 예측합니다. 이는 ground turth와 다른 정보를 추가로 얻기 때문에 BLEU에서 낮은 퍼포먼스를 보이지만 METEOR에서는 경쟁력 있는 결과를 보입니다. 이는 METEOR가 WordNet과 같은 대용어 사전을 활용하여, 비슷한 의미를 가지는 단어들의 정확도 점수를 연산함에 따라서, 새롭게 예측된 tag들이 의미론적으로 유사함을 뜻합니다. tf-idf기반 cider score는 un-frequent words에 민감하기 떄문에, 감소하는 Cider점수는 아마 새로운 정보가 추가되는 자연스러운 문장을 만드는 과정에서 변형이 있기 떄문입니다.

Figure~\ref{fig2:ab-test} shows the subjective A-vs-B test results. Each method is compared to the ground truth in terms of having more true positives (Q1) and fewer false positives (Q2). 
%We collected 480 ratings per question from 24 participants.
For the first question, compared to the baseline K2C augmentation, the proposed methods using the instructions show an overwhelmingly higher \textit{win+tie} score. This indicates the importance of music-specific instructions when utilizing LLM. In particular, ``Paraphrase'' and ``Attribute Prediction'' achieve high \textit{win} scores by incorporating new information that is different from the existing vocabulary. In the second question, all caption generation methods except ``Attribute Prediction'' show higher \textit{win+tie} scores than \textit{lose} scores.
This advocates the trustworthiness of LLM-based caption generation as it shows a similar or less false-positive rate to the ground truth. With its longest average length, ``Attribute Prediction'' turns out to be `too creative' and shows a slightly higher false-positive rate than the ground truth.

\begin{table}[!t]
\centering
\resizebox{\linewidth}{!}{
\begin{tabular}{lrrrr}
\toprule
Dataset & \# item & Duration (h) & C/A & Avg. Token \\ \midrule
\multicolumn{5}{l}{{\color[HTML]{9B9B9B} General Audio Domain}} \\
AudioCaps \cite{kim2019audiocaps} & 51k & 144.9 & 1 & 9.0\small{±N/A} \\
LAION-Audio \cite{wu2022large} & 630k & 4325.4 & 1-2 & {\color[HTML]{9B9B9B}N/A} \\
WavCaps \cite{mei2023wavcaps} & 403k & 7568.9 & 1 & 7.8\small{±N/A} \\ \midrule
\multicolumn{5}{l}{{\color[HTML]{9B9B9B} Music Domain}} \\
MusicCaps \cite{agostinelli2023musiclm} & 6k & 15.3 & 1 & 48.9\small{±17.3}\\
MuLaMCap$^{*}$ \cite{huang2023noise2music} & 393k & 1091.0 & 12 & {\color[HTML]{9B9B9B}N/A} \\
\textbf{LP-MusicCaps-MC} & 6k & 15.3 & 4 & 44.9\small{±21.3} \\
\textbf{LP-MusicCaps-MTT} & 22k & 180.3 & 4 & 24.8\small{±13.6} \\
\textbf{LP-MusicCaps-MSD} & 514k & 4283.1 & 4 & 37.3\small{±26.8}
\\ \bottomrule
\end{tabular}
}
\vspace{-3mm}
\caption{Comparison of audio-caption pair datasets. C/A stands for the number of caption per audio. {*}Although we include MuLaMCap in the table for comparison, it is not publicly accessible.}
\vspace{-3.5mm}
\label{tab:dataset}
\end{table}

\section{Dataset: LP-M\lowercase{usic}C\lowercase{aps}}
Based on the proposed pseudo caption generation method, we introduce LP-MusicCaps, an LLM-based Pseudo music caption dataset. We construct the music-to-caption pairs using three existing multi-label tag datasets and four task instructions. The data sources are MusicCaps~\cite{agostinelli2023musiclm},  
%which were used in the previous experiment, 
Magnatagtune~\cite{law2009evaluation}, and Million Song Dataset~\cite{bertin2011million} ECALS subset~\cite{doh2022toward}. We respectively refer to them as MC, MTT, and MSD. MC contains 5,521 music examples,\footnote{We only use 5495 out of the total due to the loss of 26 data samples.} each of which is labeled with 13,219 unique aspects written by music experts. MTT \cite{law2009evaluation} consists of 26k music clips from 5,223 unique songs including genre, instrument, vocal, mood, perceptual tempo, origin, and sonority features. We used the full 188 tag vocabulary and did not generate captions for tracks that do not have associated tags (decreased to 22k). MSD consists of 0.52 million 30-second clips and 1054 tag vocabulary \cite{doh2022toward}. The tag vocabulary covers various categories including genre, style, instrument, vocal, mood, theme, and culture. Each dataset uses an average of 10.7 / 3.3 / 10.2 labels per music clip for generating pseudo captions, respectively. 

Table~\ref{tab:dataset} provides a comparison of statistics between the LP-MusicCaps family and other audio-caption pair datasets. When comparing the two domains, AudioCaps~\cite{kim2019audiocaps} and MusicCaps have high-quality human annotated captions, but they have fewer captions with shorter audio duration. When comparing large-scale datasets, the music domain lacks available datasets compared to the general audio domain (such as LAION-Audio~\cite{wu2022large} and WavCaps~\cite{mei2023wavcaps}). Although MuLaMCap has an overwhelming amount of annotated captions, it is not publicly available. In contrast, LM-MusicCaps is publicly accessible and provided with various scales. LP-MusicCaps-MC has a similar caption length to manually written captions while having four times more captions per audio. LP-MusicCaps-MTT is a medium-sized dataset with audio download link, and LP-MusicCaps-MSD has the largest audio duration among various captions in the music domain caption dataset.

\begin{figure}[!t]
\centering
\includegraphics[width=\linewidth]{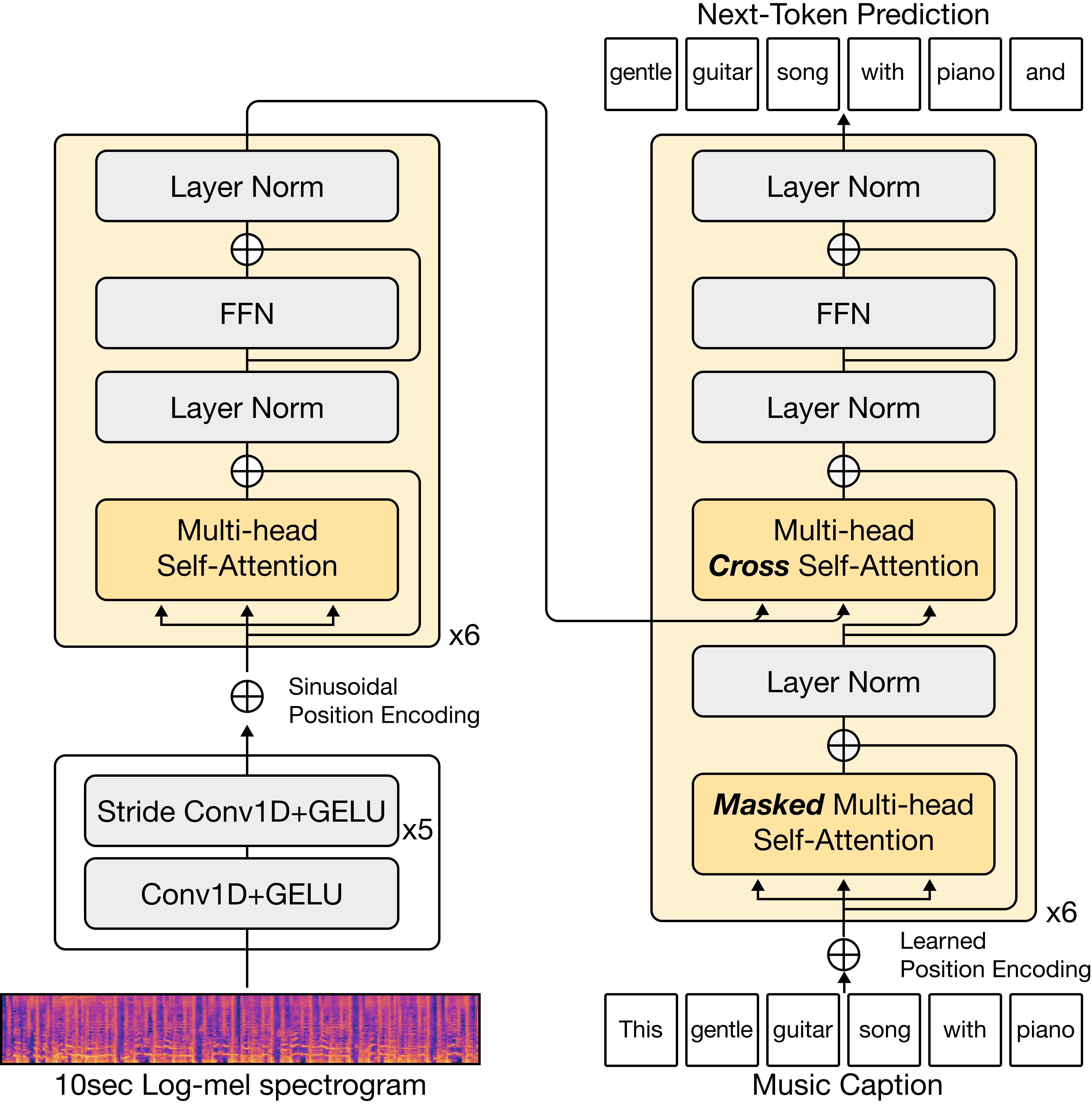}
\vspace{-7mm}
\caption{A cross-modal encoder-decoder architecture.}
\vspace{-3.5mm}
\label{fig:model}
\end{figure}

\begin{table*}[!t]
\centering
\resizebox{0.95\textwidth}{!}{
\begin{tabular}{lcccccccccccccc}
\toprule 
&  & \multicolumn{7}{c}{Supervised Metrics} &  & \multicolumn{3}{c}{Diversity Metrics} &  & Length \\
\cmidrule{3-9} \cmidrule{11-13} \cmidrule{15-15} 
Model &  & B1$\uparrow$ & B2$\uparrow$ & B3$\uparrow$ & B4$\uparrow$ & M$\uparrow$ & R-L$\uparrow$ & BERT-S$\uparrow$ &  & Vocab$\uparrow$ & $\text{Novel}_{\text{v}}\uparrow$ & $\text{Novel}_{\text{c}}\uparrow$ &  & Avg.Token \\ \midrule
\multicolumn{15}{l}{{\color[HTML]{9B9B9B} Baseline}} \\
Supervised Model &  & 28.51 & 13.76 & 7.59 & 4.79 & 20.62 & 19.22 & 87.05 &  & 2240 & 0.54 & 69.00 &  & 46.7\small{±16.5} \\ \midrule
\multicolumn{15}{l}{{\color[HTML]{9B9B9B} Zeroshot Captioning}} \\
Tag Concat~\cite{cai2020music, doh2022toward} &  & 4.33 & 0.84 & 0.26 & 0.00 & 3.10 & 2.01 & 79.30 &  & 802 & 46.38 & 100.00 &  & 23.8\small{±12.1} \\
Template~\cite{chen2022learning} &  & 7.22 & 1.58 & 0.46 & 0.00 & 5.28 & 6.81 & 81.69 &  & 787 & 45.24 & 100.00 &  & 25.8\small{±12.4} \\
K2C-Aug~\cite{wu2022large} &  & 7.67 & 2.10 & 0.49 & 0.10 & 7.94 & 11.37 & 82.99 &  & \textbf{2718} & \textbf{81.97} & 100.00 &  & 19.9\small{±7.6} \\
LP-MusicCaps \textbf{(Ours)} &  & \textbf{19.77} & \textbf{6.70} & \textbf{2.17} & \textbf{0.79} & \textbf{12.88} & \textbf{13.03} & \textbf{84.51} &  & 1686 & 47.21 & 100.00 &  & 45.3\small{±28.0} \\ \midrule
\multicolumn{15}{l}{{\color[HTML]{9B9B9B} Tansfer Learning}} \\
Tag Concat~\cite{cai2020music, doh2022toward} &  & 28.65 & 14.68 & 8.68 & 5.82 & 21.88 & 21.31 & 87.67 &  & 1637 & 3.30 & 96.07 &  & 41.8\small{±14.3} \\
Template~\cite{chen2022learning} &  & 28.41 & 14.49 & 8.59 & 5.78 & 21.88 & 21.25 & 87.72 &  & 1545 & \textbf{3.62} & \textbf{96.77} &  & 41.1\small{±13.2} \\
K2C-Aug~\cite{wu2022large} &  & \textbf{29.50} & \textbf{14.99} & 8.70 & 5.73 & 21.97 & 20.92 & 87.50 &  & \textbf{2259} & 1.42 & 84.95 &  & 44.1\small{±15.0} \\
LP-MusicCaps \textbf{(Ours)} &  & 29.09 & 14.87 & \textbf{8.93} & \textbf{6.05} & \textbf{22.39} & \textbf{21.49} & \textbf{87.78} &  & 1695 & 1.47 & 96.06 &  & 42.5\small{±14.3} \\ \bottomrule
\end{tabular}
}
\vspace{-2mm}
\caption{Music captioning results on the MusicCaps eval-set. Avg.Token stands for the average number of token per caption.}
\vspace{-3.5mm}
\end{table*}

\section{Automatic Music Captioning}\label{sec:music_captioning}
We trained a music captioning model and evaluated it under zero-shot and transfer-learning settings. This section reports the experimental results.

% We trained a music captioning model with LP-MusicCaps-MSD and evaluated it under zero-shot and transfer-learning settings. This section reports the experimental results.

\subsection{Encoder-Decoder Model}
We used a cross-modal encoder-decoder transformer architecture that has achieved outstanding results on various natural language processing tasks~\cite{lewis2019bart}, lyrics interpretation~\cite{zhang2022interpreting}, and speech recognition~\cite{radford2022robust}, as shown in Figure~\ref{fig:model}. Similar to Whisper~\cite{radford2022robust}, the encoder takes a log-mel spectrogram with six convolution layers with a filter width of 3 and the GELU~\cite{hendrycks2016gaussian} activation function. With the exception of the first layer, each convolution layer has a stride of two. The output of the convolution layers is combined with the sinusoidal position encoding and then processed by the encoder transformer blocks. Following the BART$_{\text{base}}$ architecture, our encoder and decoder both have 768 widths and 6 transformer blocks. The decoder processes tokenized text captions using transformer blocks with a multi-head attention module that includes a mask to hide future tokens for causality. The music and caption representations are fed into the cross-modal attention layer, and the head of the language model in the decoder predicts the next token autoregressively using the cross-entropy loss, formulated as: $\mathcal{L}=-\sum _{t=1}^{T} \log p_{\theta}(y_t\mid y_{1:t-1}, x)$
% $$
% \mathcal{L}=-\sum _{t=1}^{T} \log p_{\theta}(y_t\mid y_{1:t-1}, x)
% $$
where $x$ is the paired audio clip and $y_{t}$ is the ground truth token at time $t$ in a caption with length $T$. 

\subsection{Experimental Setup}
To evaluate the impact of the proposed dataset on the music captioning task, we compare a supervised model trained on the MusicCaps~\cite{agostinelli2023musiclm} training split and a pre-trained model trained on an LP-MusicCaps-MSD dataset. For the pre-trained model, we perform both a zero-shot captioning task that does not use any MusicCaps~\cite{agostinelli2023musiclm} dataset and a fine-tuning task that updates the model using MusicCaps~\cite{agostinelli2023musiclm} training split. For comparison with other pseudo caption generation methods, we report results on baseline models trained with the same architecture and amount of audio, but different pseudo captions. In addition to all the metrics we used in Section~\ref{sec:obj}, we compute \textit{Novel$_{c}$}, the percentage of generated captions that were not present in the training set~\cite{stefanini2022show}. It measures whether the captioning model is simply copying the training data or not.  

For all the experiments, the input of the encoder is a 10-second audio signal at 16~kHz sampling rate. It is converted to a log-scaled mel spectrogram with 128 mel bins, 1024-point FFT with a hann window, and a hop size of 10~ms. All models are optimized using AdamW with a learning rate of 1e-4. We use a cosine learning rate decay to zero after a warmup over the first 1000 updates. For the pre-training dataset, we use 256 batch-size and the models are trained for 32,768 updates. We adopt a balanced sampling~\cite{won2021multimodal}, which uniformly samples an anchor tag first and then selects an annotated item. For supervised and transfer learning, we use a 64 batch size, 100 epochs. We use beam search with 5 beams for the inference of all models.

% Through preliminary experiments, we discovered that the imbalanced tag distribution in MSD adversely affects model performance. 

\subsection{Results}
When comparing within zero-shot captioning models, the model trained on the proposed LP-MusicCaps dataset shows a strong performance in general. The model using tag concatenation shows the lowest performance as it fails to generate musical sentences. In case of the model using a prompt template, it demonstrates a slightly higher BERT-Score, while still exhibiting poor performance in terms of $n$-gram metrics due to its limited vocabulary. The model using K2C augmentation outperforms the other two methods but still falls short due to its lack of a musical context. In general, zero-shot models does not perform as well as the supervised baseline in most of the metrics with few exceptions. 

Among the transfer captioning models, the model with LP-MusicCaps pre-training achieves strong performance overall by winning in the BERT-Score and most of the n-gram metrics. It is noteworthy that our proposed model shows a meaningful increase in BERT-Score compared to the supervised model. This improvement is likely a result of successful semantic understanding rather than word-to-word matching. Moreover, by the improvement of Novel$_{c}$, the LP-MusicCaps model demonstrates that it can generate new captions instead of repeating the phrases in the training dataset. This advantage is observed in both the zero-shot and supervised tasks in transfer learning models.

% It is noteworthy that our proposed model shows a significant increase in CIDEr (5.95$\xrightarrow{}$10.88) compared to the supervised model. This improvement is likely due to better capturing of music-specific words that appear less frequently in the overall captions. Moreover, by the improvement of Novel$_{c}$ (68.91$\xrightarrow{}$95.99), the LP-MusicCaps model demonstrates it can generate new captions instead of repeating the training dataset. This advantage is observed in both the zero-shot and supervised tasks in transfer learning models.

\section{Conclusion}\label{sec:conclusion}
We proposed a tag-to-pseudo caption generation approach with large language models to address the data scarcity issue in automatic music captioning. We conducted a systemic evaluation of the LLM-based augmentation, resulting in the creation of the LP-MusicCaps dataset, a large-scale pseudo-music caption dataset. We also trained a music captioning model with LP-MusicCaps and showed improved generalization. Our proposed approach has the potential to significantly reduce the cost and time required for music-language dataset collection and facilitate further research in the field of connecting music and language, including representation learning, captioning, and generation. However, further collaboration with the community and human evaluation is essential to enhance the quality and accuracy of the generated captions. Additionally, we believe that exploring the use of LLMs for other topics under music information retrieval and music recommendation could lead to novel and exciting applications.

% For bibtex users:
\bibliography{ISMIRtemplate}

% \newpage 
% \appendix
% \onecolumn

% \begin{center}
% Supplementary Material for LP-MusicCaps: LLM-Based Pseudo Music Captioning
% \end{center}

% \section{LP-M\lowercase{usic}C\lowercase{aps} Dataset}
% In this paper, we release LP-MusicCaps, an LLM-based pseudo-music caption dataset.

% \noindent Example of pseudo captions:

% \section{Qualitative Evaluation}

% \section{Comparison between LLM-based Captioning and Contents-based Captioning}
\end{document}